\documentstyle[12pt,epsfig]{article}
\def\bold#1{\setbox0=\hbox{$#1$}%
     \kern-.025em\copy0\kern-\wd0
     \kern.05em\copy0\kern-\wd0
     \kern-.025em\raise.0433em\box0 }

\newcommand{\ifb}{\,\mbox{fb}^{-1}}
\newcommand{\gev}{\,\mbox{GeV}}
\newcommand{\tev}{\,\mbox{TeV}}
\newcommand{\nz}[1]{\tilde{\chi}_{#1}^0}
\newcommand{\cm}[1]{\tilde{\chi}_{#1}^-}

\newcommand{\cp}[1]{\tilde{\chi}_{#1}^+}
\newcommand{\mne}[1]{m_{\tilde{\chi}^0_{\SP {#1}}}}
\providecommand{\SP}{\scriptscriptstyle}

\def\thm{\theta_\mu}
\def\tha{\theta_A}
\def\th#1{\phi_#1}
\def\tb{\tan\beta}
\def\ga{\mathrel{\raise.3ex\hbox{$>$\kern-.75em\lower1ex\hbox{$\sim$}}}}
\def\la{\mathrel{\raise.3ex\hbox{$<$\kern-.75em\lower1ex\hbox{$\sim$}}}}

\def\mcha{m_{\chi^{\pm}}}

\def\m12{m_{1\!/2}}

\def\ohsq{\Omega_{\widetilde\chi}\, h^2}
\def\bino{\widetilde B}

\begin{document}
\begin{titlepage}
\pagestyle{empty}
\baselineskip=21pt
\rightline{hep-ph/0101106}
\rightline{MADPH-00-1207}
\rightline{TPI-00164}
\rightline{UPR-916-T}
\vskip 1in
\begin{center}
{\large{\bf
CP-Violating Phases in SUSY, Electric Dipole Moments, and Linear Colliders
}}
\end{center}
\begin{center}
  \vskip 0.2in {\bf V. Barger, 
T. Falk,\footnote{Current address: Theoretical Physics Institute, School of Physics and Astronomy, University of Minnesota, Minneapolis, MN 55455} T. Han, J. Jiang, \\
T. Li,\footnote{Current address: Department of Physics and 
Astronomy, University of Pennsylvania, Philadelphia, PA 19104}
 and T. Plehn }\\
  \vskip 0.1in {\it Department of Physics, University of Wisconsin,
    Madison, WI~53706, USA} \vskip 0.2in {\bf Abstract}
\end{center}
\baselineskip=18pt \noindent We reexamine large CP-violating phases in
the general Minimal Supersymmetric Standard Model, as well as more
restricted models.  We perform a detailed scan over parameter space to
find solutions which satisfy the current experimental limits on the
electric dipole moments of the electron, neutron and $^{199}$Hg atom,
exploring the allowed configurations of phases and masses, and we
attempt to quantify the level of tuning of the parameters
necessary to populate the regions of cancellations.  We then consider
the measurement of CP-violating phases at a future linear collider.
We find that measurements of chargino and neutralino masses and
production cross-sections allow for a determination of 
$\phi_1$(the phase of $M_1$) to
a precision of $\pi/30$, while the EDM constraints require that
$\theta_\mu$ be too small to be measured.  Using the EDM constraints we
find that the CP-even model parameters and the phase $\phi_1$ can be
determined at a Linear Collider with $400 \gev$ c.m.\ energy. 
As long as some
information on the size of $|\mu|$ is included in the observables, a
measurement of $\phi_1$ is guaranteed for $\phi_1 > \pi/10$. To
unambiguously identify CP violation, we construct CP-odd kinematical
variables at a linear collider. However, the CP asymmetries are rather
small, typically about $0.1-1.5\%$, and it will be challenging to
experimentally observe the predicted
asymmetries.
\end{titlepage}
\baselineskip=18pt
\setcounter{footnote}{0}

\section{Introduction}

In the last two years, there has been a resurgence of interest in 
CP-violating phases in supersymmetric theories.  Although in many cases
the size of the CP-violating phases is strongly constrained by limits
on the electric dipole moments (EDMs) of the electron, neutron and
$^{199}$Hg atom, possible cancellations between different
contributions to the EDMs can significantly weaken the upper bounds on
the phases \cite{FO}, even potentially allowing phases of ${\cal
  O}(1)$ \cite{IN,Kane}, typically in small, often finely tuned
regions.  This observation has led to an explosion of papers exploring
the consequences of the presence of large phases, many papers finding
some regions of parameter space where large phases are of consequence
to the system under study, but few papers simultaneously imposing the
(still severe) constraints from EDMs.  One of the most interesting
questions is whether one will be able to measure the phases in
chargino and neutralino production at a future linear collider
\cite{mad,cpodd}.  Even if one expects that weakly
interacting supersymmetric partners are going to be found at the
Tevatron and at the LHC~\cite{report,hadron}, the model parameters
have to be determined in detail in a precision experiment. A future
high luminosity Linear Collider is the optimal tool to extend the
LEP Standard Model precision measurements to other models. The 
CP-conserving set of underlying MSSM parameters can be easily
determined from physical masses alone, cross sections and
asymmetries~\cite{cpeven}.  The effects of CP-violating phases on
neutralino and chargino observables can be considerably smaller and therefore
require a more ambitious collider energy and
luminosity~\cite{mad,cpodd}. 

In this paper we reexamine this question, taking into
account the limits from EDM experiments, including the recent improved limits 
on the $^{199}$Hg EDM \cite{mEDM}.  We perform a
detailed scan of parameter space in the general (non-unified) MSSM,
including phases and masses, to find the regions satisfying the EDM
constraints.  Using these constraints we show how the phase $\phi_1$ 
can be extracted from a set of CP-even variables (masses and total 
cross sections) at a future Linear Collider. Compared to the previous 
analysis~\cite{mad} we can reduce the requirements on the energy
as well as on the luminosity of the Linear Collider. Moreover, 
we show how  one could use an extended set of parameters, which 
allows for an independent extraction of $|\mu|$, 
and further relax the requirement on the Collider design parameters.   
Finally, we construct CP-odd variables at $e^+e^-$ linear colliders.
We find that the CP asymmetry is typically about $0.1-1.5\%$. 

\section{CP Violation and Electric Dipole Moments\label{sec:edms}}

The Minimal Supersymmetric Standard Model (MSSM) contains several
sources of CP violation not present in the Standard Model.  In the
most general flavor non-preserving SUSY model, there are over 40 new
complex phases \cite{ds}, although most of the new parameters are very
strongly constrained by limits on flavor violating processes.  We
assume a Peccei-Quinn symmetry \cite{Peccei:1977hh} which sets the coefficient $\bar\theta$ of the
$\tilde G G$ term to zero, up to small corrections coming from
higher dimensional CP-violating operators, which shift the minimum of
the axion potential \cite{bigi}.  In the MSSM, new complex phases arise in the
Higgs mixing mass $\mu$, as well as in the soft SUSY-breaking terms in
the Lagrangian: the trilinear scalar mixing masses $A_i$, the bilinear
Higgs mixing parameter $B$, and the three gaugino mass parameters $M_i$.
Not all of these phases are physical, however, and some or most may be
removed by field redefinitions, depending on the model.  In mSUGRA,
for example, the phases $\phi_i$ of all the $M_i$ may be removed by an
R-rotation\footnote{Under an R-rotation of the fields, 
the $A_i, B$ and $M_i$ effectively pick up a common
phase, while the other soft SUSY breaking masses and $\mu$ are left unchanged.} at the unification scale
$M_X$, the $A_i$ are united to a common $A_0$ at $M_X$, and a rotation
of one of the Higgs fields can be used to set $\theta_\mu+\theta_B=0$,
so that the Higgs vevs are real.  This leaves only two physical phases
in mSUGRA, which can be taken to be $\theta_{A_0}$ and $\theta_\mu$.

In more general models, where the $M_i$ do not unify, only one of the
gaugino phases can be rotated away, which we take to be the phase of
$M_2$.  In our more general analyses, we will also take independent
phases in the trilinear parameters $A_e, A_u, A_d$ and $A_t$.  This
leaves us with seven physical phases in our most general model.  We
emphasize that these are the phases present {\it after} phases have
been removed in the field redefinitions described above.  Therefore
when we constrain below the phase of the $\mu$ parameter to be very
small, it is in fact some combination of the phases in the original
parameterization which is restricted.  This is particularly
important in models where there is a correlation between the phase of
$B$ and the other phases in the model, such as in mSUGRA, as we
discuss further below.

It is well known that the additional sources of CP violation in
supersymmetric models can contribute to the electric dipole moments of 
the neutron and electron \cite{dn,ko} and mercury atom \cite{fopr}.
The very tight experimental limits on these quantities \cite{nEDM,eEDM,mEDM} 
\begin{eqnarray}
d_n &<& 1.1\times10^{-25}\;e\cdot{\rm cm}\label{eq:elim}\\
d_e &<& 4\times10^{-27}\;e\cdot{\rm cm}\\
d_{\rm Hg} &<& 2\times10^{-28}\;e\cdot{\rm cm}\label{eq:hglim}
\end{eqnarray}
then impose severe constraints on the CP-violating phases in SUSY models.
To suppress the EDMs, either large scalar masses ( $>1$ TeV) or small
phases (of the order $10^{-3}$, when all SUSY masses are of order 100
GeV) are typically required.  However, as pointed out in
\cite{FOS,FO,FO1}, such large scalar masses are cosmologically
problematic, and the addition of cosmological constraints to the mix
implies that (some of) the CP-violating phases are constrained to be
quite small when the LSP is a dominantly $\bino$-type neutralino.

\begin{figure}[h,t]
  \begin{center}
  \epsfig{file=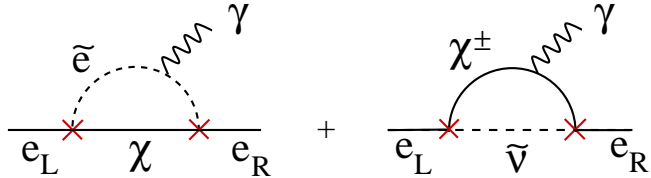, height=1.0in}     
    \caption{Diagrams contributing to the electric dipole moment of
the electron.  The SUSY phases enter at the crossed vertices.}
    \label{fig:ediags}
  \end{center}
\end{figure}

The electron EDM receives contributions from chargino and neutralino
exchange diagrams, shown in Fig.~\ref{fig:ediags}.  The full
expressions for the dependence of the induced electron EDM on the SUSY
masses and phases can be found in \cite{ko,IN}.  The dominant
contribution is typically from the chargino exchange diagram and is
proportional to $\sin\thm$.  Thus the primary constraint coming from
the electron EDM limits is an upper limit on the phase of $\mu$. 
 The subdominant neutralino exchange piece has a more complicated
 dependence on the SUSY phases, including both pieces proportional to
 $\sin\thm$ and $\sin\gamma_e$, where $\gamma_e=\arg(A_e + \mu^*\tb)$.
 Cancellations can occur between the neutralino chargino exchange
 contributions, and this serves to weaken the absolute limits on the
 SUSY phases, although $\thm$ of ${\cal O}(1)$ requires either severe
 fine-tuning of parameters or a very heavy spectrum, as we will see below.

The neutron EDM is considerably more complicated, and until recently,
the computation \cite{ko,IN} of the neutron EDM induced by SUSY phases
has been plagued by very large theoretical uncertainties.  The SUSY
phases contribute both to the EDMs and color EDMs (cEDMs) of the
quarks, and in the last year, the contribution to the neutron EDM both 
from the induced $\bar\theta$ 
due to the color EDMs of the quarks
\cite{PRtheta} and from the quark EDMs and cEDMs themselves \cite{PRnedm}
have been reliably calculated using QCD sum rules, allowing a 
reduction in the theoretical uncertainty.  The overall effect is to
reduce slightly the predicted neutron EDM, and with smaller error
bars.
 \begin{figure}[h,t]
  \begin{center}
  \epsfig{file=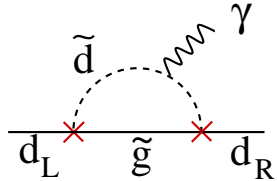, height=1.0in}     
    \caption{Diagram contributing to the electric dipole moment of the down quark. The SUSY phases enter at the crossed vertices.}
    \label{fig:gluino}
  \end{center} 
\end{figure} 
The quark EDMs receive contributions from the chargino and neutralino
exchange diagrams of Fig.~\ref{fig:ediags} (where the photon can now
also connect to the sfermion line in the chargino diagram), as well as
from the gluino exchange diagram of Fig.~\ref{fig:gluino}.  The gluino
exchange contribution to the $d$-quark EDM is proportional to
$\gamma_d=\arg(A_d + \mu^*\tb)$ (for the $u$-quark take
$\tb\rightarrow\cot\beta$).  There are also significant contributions
to the neutron EDM from the color dipole moments of the quarks, which
have the same dependence on the phases at the quark EDMs, and are of
the same order as the contribution from the induced $\bar\theta$ term
described above.  The neutron EDM written in terms of the quark EDMs
and cEDMs (which can be found in \cite{IN,ko}) is given in
\cite{PRnedm}.  Lastly, there is also a small contribution from the
three-gluon operator $O_G = - {1 \over 6} f^{abc}G_aG_b{\tilde G_c}$
\cite{weinberg}. Two-loop contributions to the EDMs may be significant
at very large values of $\tan\beta$~\cite{pilaftsis}. For the values
of $\tan\beta$ we consider, the effect of higher order contributions
will be to very slightly shift the allowed regions of parameter space.

For the $^{199}$Hg atom, the electric screening of the electric dipole
moments of the atom's constituents is violated by the finite size of
the nucleus and can be conveniently expressed by the Schiff moment
$S$, which parameterizes the effective CP-odd interaction between the
electron and nucleus of spin $I$.  The Schiff moment, in turn, can be
induced either through the Schiff moment of the valence nucleons or
through the breaking of time reversal invariance in the
nucleon-nucleon interaction, the latter being enhanced by the
collective effects in the nucleus, and hence is dominant.  The largest
contributions to the Schiff moment, and hence to the EDM of mercury,
are through the color EDMs of the quarks, and the expression for
$d_{\rm Hg}$ in terms of the quark cEDMs can be found in \cite{fopr}.
We emphasize that the mercury, electron and neutron EDMs all depend on
different combinations of phases, and so simultaneously imposing all
three EDM constraints excludes a much greater portion of the
CP-violating SUSY parameter space than from imposing any single
constraint.


Again, the EDMs generated by the SUSY phases are sufficiently small if
either 1)~the phases are very small ($\la 10^{-2}-10^{-3}$), or 2)~the SUSY
masses are very large (${\cal O}$~(a few TeV)), or 3)~there are large
cancellations between different contributions to the EDMs, or by a
combination of these effects.  It is condition 3), large
cancellations between different contributions to the electric dipole
moments, which has spurred the greatest interest recently, since it
ostensibly allows for both large amounts of CP violation and a
spectrum which is phenomenologically relevant.  Such cancellations are
not easy to achieve, however.  If the phases are ${\cal O}(1)$ and the
SUSY masses are in the 100 GeV range, then the EDM limits will be
violated by several orders of magnitude unless very delicate
cancellations between the various contributions exist.  Further, the
parameters must be tuned so that similarly delicate cancellations
occur for electron {\it and} neutron {\it and} Hg EDMs.  

In some models it is impossible to achieve sufficient cancellations in all three EDMs to
permit some phases to be ${\cal O}(1)$, but even so, the
effect of cancellations may still be significant.
For example, cancellations do occur naturally
in mSUGRA models over significant regions of parameter space
\cite{FO,IN,FO1,fopr}, including in the body of the cosmologically
allowed region with $\m12={\cal O} (100-400\;{\rm GeV})$.  With a
$\bino$-type neutralino LSP, large sparticle masses cannot be 
invoked to
permit large phases, due to limits on the LSP relic abundance
\cite{FOS,FO,FO1}, and so the phases, in particular $\thm$, are
severely limited by the EDMs.  The presence of cancellations relaxes
the constraints on the phases, but the limit on $\theta_\mu$ remains
small, $\theta_\mu\la\pi/10$, unless the sfermions (including
selectrons) are heavier than ${\cal O}$~(1~TeV).  This upper
bound on $\thm$ is over an order of magnitude less restrictive 
than what would
find in the absence of any cancellations, however, and thus the effect
of cancellations is quite significant.  We will discuss the role of
cancellations in the MSSM in more detail in Section~\ref{sec:mssm15}.

Minimal SUGRA is a particularly restrictive model in that 
there are only two new physical
phases.   In more general models there are more phases, and studies have
been made to examine the new cancellations which the presence of
additional phases allow, particularly in models without gaugino mass
unification, where the gaugino masses $M_i$ can have independent
phases (see e.g. \cite{Kane,cancel}).  The possibility of
having both ${\cal O}(1)$ phases and reasonable sparticle masses in
these models has inspired a remarkable  number of recent papers
exploring the consequences (phenomenological and  cosmological) of
new large sources of CP violation in SUSY models.    Most of
these analyses take the possibility of cancellations as 
{\it carte blanche} to consider all sets of masses and phases, 
without actually imposing the
rigorous constraints on the SUSY parameters from the 
electric dipole moments.
In the next section we will examine the size of the phases one may
reasonably expect to satisfy the EDM constraints, and for what
sparticle masses,  in both mSUGRA and in two more general models
without gaugino mass unification.  In particular, we will study the
level of tuning required to obtain large phases.

\section{EDM Analysis}
We begin by studying the constraints imposed by the EDMs in three
different models: one mSUGRA-inspired model with two physical phases,
and two models without gaugino mass unification, which have seven
independent phases each and 15(23) total free parameters, including
masses.  We have done large Monte Carlo studies for each case, 
evaluating
roughly 800,000 parameter sets in mSUGRA and 300 and 600 million sets 
in each
of the two more general models respectively, and 
studied the configurations which satisfy
the experimental limits on the EDMs.  Table~\ref{tab:mc} displays
the number of parameter sets studied and total number of points
satisfying the EDM constraints (\ref{eq:elim})-(\ref{eq:hglim}).
\begin{table}
\begin{center}
\begin{tabular}{|l||c|c|c|}
\hline
 Model       &  Points run & Points satisfying EDMs & fraction satisfying EDMs \\
 \hline\hline
mSUGRA&  $8\times 10^5$  &  10,000    & $1.2\times 10^{-2}$        \\
15 parameter MSSM&  $3\times 10^8$ &  23,650    & $7.1\times 10^{-5}$  \\
23 parameter MSSM& $6\times 10^8$    &  13,700    & $2.1\times 10^{-5}$   \\
\hline
\end{tabular}
\caption{ Monte Carlo studies of electric dipole moments in three models.\label{tab:mc}}
\end{center}
\end{table}

\subsection{mSUGRA-inspired Model}
After performing the field redefinitions described in
Section~\ref{sec:edms}, the mSUGRA-like model is specified by 6
parameters: three masses ($m_0, \m12, $ and $A_0$), two phases ($\thm(M_X)$ 
and $\tha (M_X)$), and $\tan\beta$.  Throughout we take 
$m_{\rm top}=175 $ GeV.  In mSUGRA, once the gaugino, soft scalar masses, $A$
and $B$-terms and phases are given at $M_X$, they can be evolved
using the renormalization group equations (RGE)
to the electroweak scale.  
As in common practice, we use the one-loop RGEs for
the masses and two-loop RGEs for the gauge and Yukawa couplings
\cite{ikkt}. The structure of the equations for the gauge couplings,
gaugino masses and the diagonal elements of the sfermion masses are
such that they are entirely real.  The evolutions of the $A_i$,
however, are more complicated, as the $A_i$ pick up both real and
imaginary contributions.  For example, the evolution of $A_t$ is given
by
\begin{eqnarray}
  {dA_t \over dt} = {1\over{8\pi^2}}\left(-{16\over3}\,g_3^2\, M_3
    -3g_2^2\, M_2-{13\over9}\, g_1^2\, M_1 + h_b^2 A_b + 6 h_t^2 A_t\right).
\end{eqnarray}
Thus, $A_t$ receives real contributions proportional to the
gaugino masses $M_i$ and complex contributions from the heavy generation $A_i$, multiplied with the respective Yukawa coupling $h_i$.
Since the coefficients of the $M_i$ are flavor dependent and the
coefficients of the $h_f^2 A_f$ terms are generation dependent, the
phases (and magnitudes) of the $A_i$ must therefore be run separately.
At one loop, the evolution equation for $\mu$ is given by
\begin{eqnarray}
  {d\mu \over dt} = {\mu\over{16\pi^2}}\left(-3g_2^2-g_1^2+h_\tau^2+
    3h_b^2+3h_t^2\right),
\end{eqnarray}
and the phase of $\mu$ does not run.  Finally, the $B$ parameter 
evolves as
\begin{eqnarray}
{dB \over dt} = {1\over{8\pi^2}}\left(
    -3g_2^2\, M_2-g_1^2\, M_1 + h_\tau^2 A_\tau + 3 h_b^2 A_b + 3 h_t^2 A_t\right).
\label{brge}
\end{eqnarray}
After evolving the parameters to the weak scale, the phase of 
Higgs superfield
$H_2$ (which gives mass to up-type fermions) can be rotated in
such a way as to ensure real expectation values for the Higgs
scalars. The rotation changes the phase of $H_2$ by an amount 
$-(\thm+ \theta_B)$. Not only is the phase of $\mu$ now fixed at 
$\thm =-\theta_B$, but also the initial phase of $\mu$ is physically
irrelevant as it is canceled by the rotation.  As emphasized in
\cite{gw,aad}, a large phase in $A$ will induce a phase in $B$ (Eq.
(\ref{brge})), and hence in $\mu$ , after the vevs are made real.
Therefore even if $\mu$ and $B$ are both real at $M_X$, if $\tha$ is
large, the value of $\thm(M_Z)$ contributing to the EDMs may be large
(of course this is completely equivalent to keeping the Higgs vevs
complex and $\mu$ real).  Since $B$ is a free parameter, which is
typically determined by the conditions of correct electroweak
symmetry breaking, $B(M_X)$ can be chosen so that $B\mu(M_Z)$ is
nearly real, yielding a small $\thm(M_Z)$ after the Higgs rotation;
however, this can involve a significant fine-tuning if $\tha$ is large
\cite{aad}.  In practice, this tuning is typically not worse than at
the level of 10\% \cite{FO1}, but it must be emphasized that this is a
tuning {\it over-and-above} the tuning discussed below.
Alternatively, if $\theta_B(M_X)$ is taken real \cite{gw} and $\tha$ is large, or if
$\theta_B(M_X)$ is arbitrary, then the additional fine tuning lies
in adjusting the original $\thm^0$ close to $-\theta_B(M_Z)$.

To study the phases permitted by the current
experimental limits (\ref{eq:elim})-(\ref{eq:hglim}) on the neutron,
electron and $^{199}$Hg EDMs, we perform a Monte Carlo studies,
 sampling the 6
model parameters in the following ranges: 
\begin{center} \begin{tabular}{rcl}
 0    $\;\;\le$ &$\thm, \tha$ &$\le\;\;$   $2\pi$\\
50 {\rm GeV}    $\;\;\le$& $\m12$&   $\le\;\;$ 1 {\rm TeV}\\
0 {\rm GeV}    $\;\;\le$& $A_0$ &  $\le\;\;$ 1 {\rm TeV}\\
0 {\rm GeV}    $\;\;\le$& $m_0$ &  $\le\;\;$ 1 {\rm TeV}\\
2    $\;\;\le$&$\tb$&  $\le\;\;$ 10
\end{tabular} \end{center}
and keeping parameter sets satisfying
(\ref{eq:elim})-(\ref{eq:hglim}).  Limits from particle searches have
not been imposed.  However, the lower limit on $\m12$ was chosen to
remove most of the area excluded by the current LEP chargino mass
limit of 103 GeV\cite{susylim}.  Most of the solutions 
with $\tb\la3$ are excluded by the Higgs searches at 
LEP\cite{susylim}.  Cosmological limits
on the relic abundance of the LSP neutralino, which tend to exclude
regions with $m_0\ga200$ GeV,  have also not been included, we
will comment on these regions later.  

\begin{figure}[b]
\begin{center}
\epsfig{file=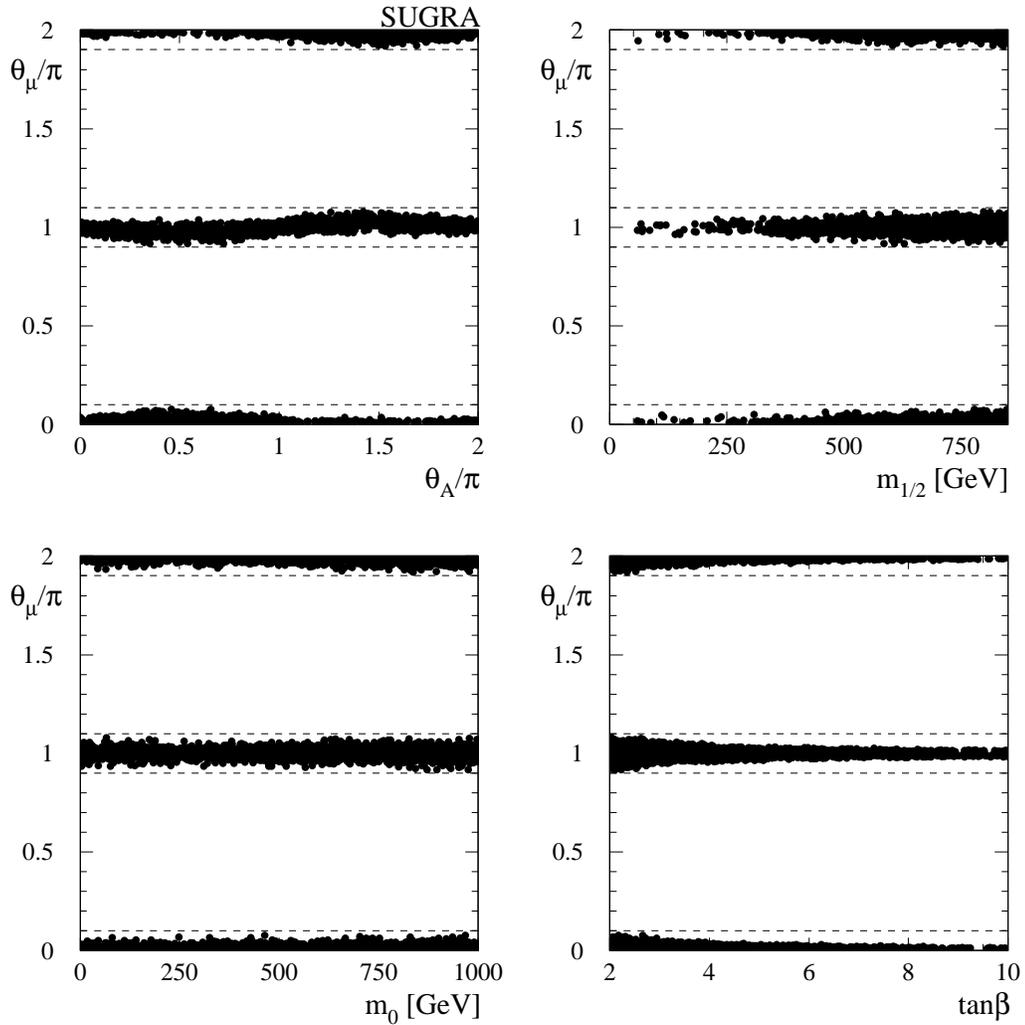,width=14cm}
\end{center} 
\vspace*{-8mm}
 \caption{mSUGRA solutions satisfying the experimental limits on the electron, neutron and $^{199}$Hg EDMs.}
    \label{fig:sugra1}
\end{figure}

The results for our Monte Carlo scan of 800,000 mSUGRA parameter sets
are summarized in Fig.~\ref{fig:sugra1}, where we display the 10,000
sets\footnote{To reduce the file size in this and following plots, we
  do not display points lying underneath covered regions, particularly
  at low $\thm$.  This
  produces no visible effect on the plots.} satisfying the three EDM
constraints (\ref{eq:elim})-(\ref{eq:hglim}).  Fig.~\ref{fig:sugra1}a
shows the allowed configurations of $\{\thm,\tha\}$.  The first
obvious point to note is that the range of $\thm$ is severely limited,
with no events in this sample having $|\thm|> 0.1\pi$ (mod $\pi$),
while the phase $\tha$ can take any value.  This is due to the fact
that the chargino contribution to the electron EDM is typically
dominant and depends only on $\thm$; therefore for large $\thm$, 
the neutralino exchange piece cannot
provide sufficient cancellation, regardless of the values of $\tha$.
In Figs.~\ref{fig:sugra1}b-\ref{fig:sugra1}d, we display the value of
$\thm$ versus $\m12$, $m_0$ and $\tb$, respectively.  We see that
there are many more solutions for large $\m12>500$ GeV, 
where the heavy spectrum reduces the individual contributions
to the EDMs.  We note,
however, that almost all such parameter sets lead to a neutralino relic
abundance $\ohsq>0.3$ \cite{fopr}, implying a universe younger than 12
billion years, in contradiction to observational evidence .  In
Fig.~\ref{fig:sugra1}d, we see that most of the large $\thm$ solutions
occur also for $\tb\la3$, which in mSUGRA yield a Higgs scalar too
light to be compatible with the negative results from Higgs searches
at LEP \cite{susylim}.  Indeed, if we consider only those parameter sets
with $\m12<500\gev$, $m_0<200\gev$ and $\tb>3$, we find no
solutions with $\thm > \pi/20$.

\subsection{15-parameter MSSM}\label{sec:mssm15}
We next consider a more general model which has additional independent
phases, and which therefore has greater opportunity for
cancellations.  We no longer require gaugino mass unification.
Therefore, one of the three gaugino masses may still be made real by
an R-rotation, but the other two, which we take to be $M_1$ and $M_3$,
may be complex, with phases $\th1$ and $\th3$ respectively.  We
additionally allow independent phases in $A_d$, $A_u$, $A_e$ and
$A_t$, and along with $\thm$, this gives 7 independent phases.  The other
8 parameters are $\tb$, plus the masses: $|M_i|, i=1\ldots3,$ a common
$|A_i|=|A|, i=e,d,u,t , |\mu|, $ and the sfermion mass parameters
$m^2_{\tilde e_R}, m^2_{\tilde u_R}$.  We fix the remaining 
masses by the
approximate relations $m^2_{\tilde d_R}=m^2_{\tilde u_R}, 
m^2_{\tilde e_L}=m^2_{\tilde e_R}+0.6
M^2_2,$ and $m^2_{\tilde q_L}=m^2_{\tilde q_R}+0.5 M^2_2$. 
We then sample the regions
\begin{center} \begin{tabular}{rcl}
 0    $\;\;\le$ &$\thm, \phi_i,{\tha}_i$ &$\le\;\;$   $2\pi$\\
100 {\rm GeV}    $\;\;\le$& $\mu$&   $\le\;\;$ 1 {\rm TeV}\\
100 {\rm GeV}    $\;\;\le$& $2M_1, M_2,M_3$&   $\le\;\;$ 1 {\rm TeV}\\
0 {\rm GeV}    $\;\;\le$& $|A|$ &  $\le\;\;$ 1 {\rm TeV}\\
0 {\rm GeV}    $\;\;\le$& $m_{\tilde e_R},m_{\tilde u_R}$ &  $\le\;\;$ 1 {\rm TeV}\\
2    $\;\;\le$&$\tb$&  $\le\;\;$ 10
\end{tabular} \end{center}
and look for parameter sets satisfying
(\ref{eq:elim})-(\ref{eq:hglim}).  We cut out the (few) remaining sets
with $\mcha<103\gev$, and we flag the sets with $m_h<113\gev$, as
discussed below.

\begin{figure}[htbp]
\begin{center}
  \epsfig{file=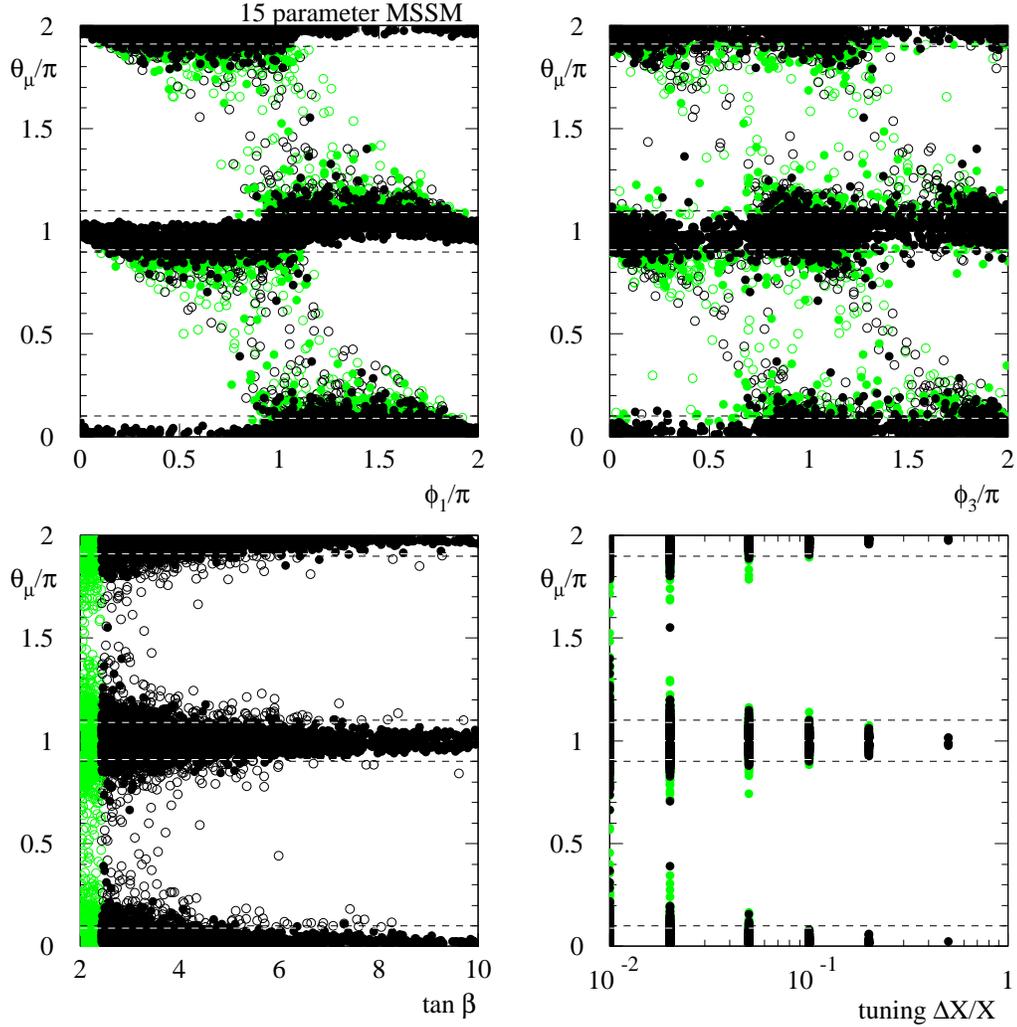,width=14cm} 
\end{center}
\vspace*{-8mm}
  \caption{Parameter sets in the 15-parameter MSSM satisfying the
    experimental limits on the electron, neutron and $^{199}$Hg
    EDMs.  Open circles  suffer from 
    parameter tuning $\Delta X/X$ worse than 1\% (see
    the text).  Light (green) dots correspond to configurations with a
    light Higgs $m_h<113\gev$.  The tuning parameter $\Delta X/X$ is
    defined in the text, and corresponds to the maximum variation that
    the point survives.}
    \label{fig:mssm151}
\end{figure}

\begin{figure}[b]
\begin{center}
\epsfig{file=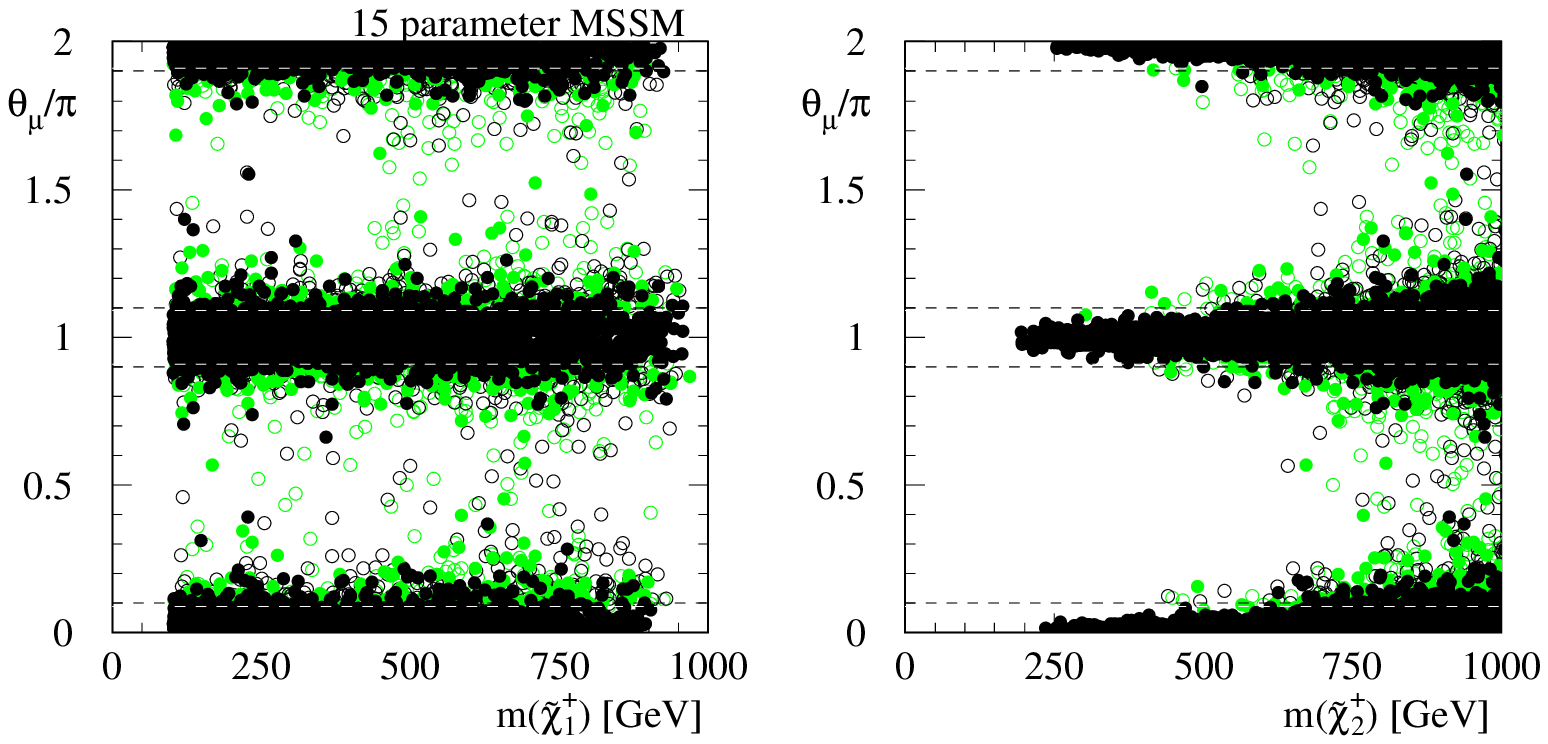,width=14cm}
\end{center}
\vspace*{-8mm}
  \caption{For the 15-parameter MSSM scan, the dependence of the
    solutions on the light and heavy chargino masses. Open circles 
    suffer from parameter
    tuning $\Delta X/X$ worse than 1\% (see the text).  Light (green)
    dots correspond to configurations with a light Higgs
    $m_h<113\gev$.}
    \label{fig:mssm153}
\end{figure}

\begin{figure}[b]
\begin{center}
\epsfig{file=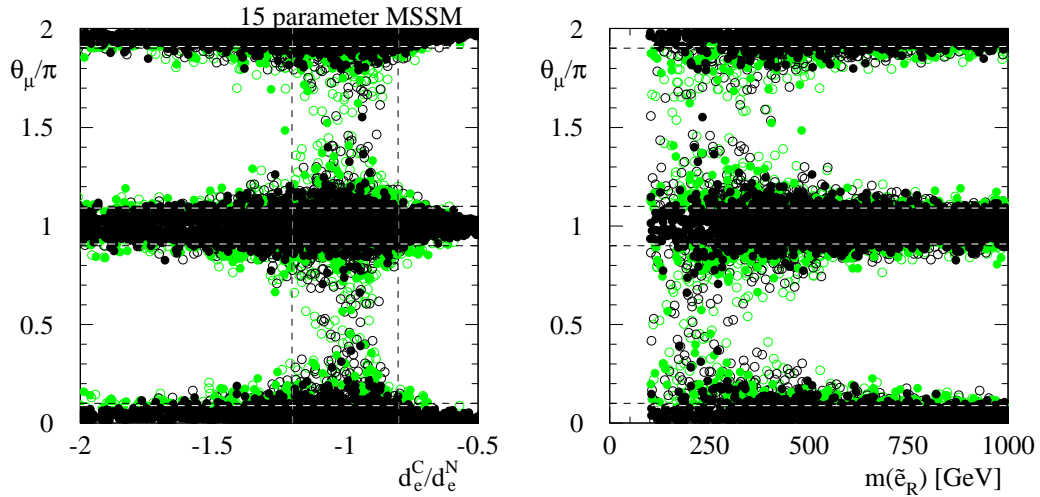,width=14cm}
\end{center}
\vspace*{-8mm}
  \caption{For the 15-parameter MSSM scan, the effect of cancellations
    on the allowed range of $\thm$. Open circles suffer from parameter
    tuning $\Delta X/X$ worse than 1\% (see the text).  Light (green)
    dots correspond to configurations with a light Higgs
    $m_h<113\gev$. }
    \label{fig:mssm152}
\end{figure}

Again, because of the additional possibilities for cancellations in
this parameterization, we expect solutions with larger values of the
phases which still satisfy the EDM constraints
(\ref{eq:elim})-(\ref{eq:hglim}).  This is borne out in
Figs.~\ref{fig:mssm151}-\ref{fig:mssm152}, where we display the
results for our Monte Carlo scan of roughly 300 million points in the
15-parameter MSSM.  In the top two panels of Fig.~\ref{fig:mssm151} we
display the 23,500 sets allowed by the EDMs, in the $\{\thm,\th1\},$
and $\{\thm,\th3\}$ planes, respectively.  Although by far the
greatest events of solutions have $\thm\la\pi/10$, there is now a
visible swath that extends to large values of $\thm$.  The range of
$\th1$ and $\th3$ are unconstrained, as are the ranges of the
$\theta_{A_i}$ (not displayed).  The presence of the large $\thm$
solutions relies on having large sparticle masses, significant
cancellations between contributions to the EDMs, or small $\tb$, and
very typically a combination of the above.

In Fig.~\ref{fig:mssm151} we display $\thm$ versus $\tb$ for the
parameter sets, and we see clearly the weakened constraint on $\thm$
at low $\tb$.  However, most of the low $\tb$ solutions are actually
experimentally excluded by the unsuccessful Higgs searches at LEP2.
To be conservative, we take the stop soft masses to be independent
parameters and set them to 1 TeV, take $m_A=1~\tev$, and we compute
the light Higgs mass as a function of $\tb$ in the maximal mixing
scenario using Ref.~\cite{hhw}.  The light (green) points in all four
panels of Fig.~\ref{fig:mssm151} yield $m_h<113\gev$ (corresponding to
$\tb\la2.4$). Hence, they would require large phases and further tuning in the Higgs sector not to be excluded by LEP~\cite{susylim}. In
Fig.~\ref{fig:mssm153}, we display $\thm$ as a function of the light
($\tilde\chi^+_1$) and heavy ($\tilde\chi^+_2$) chargino masses.  The
chargino exchange contribution to the fermion EDMs and cEDMs
\cite{ko,IN} is suppressed both for highly split charginos and for a
heavy chargino spectrum.  We see that if neither chargino is heavy
($\mcha\la500\gev$), $\thm$ is strongly constrained, whereas if at
least one of the charginos is heavy, $\thm$ can potentially be large.
In particular, having a spectrum with only one light chargino does not
by itself  forbid $\thm>\pi/10$.

The significance of cancellations on the allowed range of $\thm$ is
seen in Fig.~\ref{fig:mssm152}.  In the first panel we display the
value of $\thm$ versus the ratio the the chargino ($d_e^C$) and
neutralino ($d_e^N$) exchange contributions to the electron EDM.  We
see the clustering of the large $\thm$ solutions near to the
cancellation point $d_e^C/d_e^N=-1$.  The vertical lines correspond to
an electron EDM a factor of 5 smaller than the larger of $ d_e^C$ and $d_e^N$,
 due to
cancellations.  This demonstrates that the sparticles must be much
heavier than those allowed by our range of parameters ($> 1 \tev$) in
order for the EDM constraints to be satisfied for large $\thm$ in the
absence of any cancellations.  In practice, the large $\thm$ solutions
have both some sparticle masses near the upper end of their  ranges
{\it and} a large degree of cancellations.  In the second panel of
Fig.~\ref{fig:mssm152}, we display $\thm$ as a function of the
sfermion mass $m_{\tilde {e_{\SP R}}}$.  Perhaps counter-intuitively,
the large $\thm$ solutions tend to lie at lower $m_{\tilde e_{\SP
  R}}$, rather than near the upper end of their range.  However, this
is again due to the necessity for cancellations: $m_{\tilde
  \ell}\sim1\tev$ is not large enough to sufficiently turn off the
SUSY contributions to the electron EDM for large $\thm$, and the
necessary cancellations only occur for smaller slepton masses.

One goal of this paper is to study the extent to which the large phase
solutions require a significant tuning of the model parameters in
order to fall into the regions of EDM cancellations.  The tuning
measure we employ is simple but intuitive.  For every parameter set we
find which satisfies the EDM constraints, we perform the following
analysis.  We begin by varying all the input parameters (one at a
time) by $\Delta X/X=\pm 0.5\%$ and see if the EDM limits are still
satisfied for all the test parameter sets.  If they are, we then try
varying all the input parameters by $\pm 1\%$, and so on, until we
find the smallest percentage change for which the configuration
violates one of the EDM bounds.  This gives a sense of the local
``size'' of the allowed parameter region.  The results for the
15-parameter MSSM scan are displayed in the final panel of
Fig.~\ref{fig:mssm151}, where we have stepped through 0.5\%, 1\%, 2\%,
5\%, 10\%, 20\% and 50\% changes, modulo $\pi$ in the case of phases,
and the last percentage change to the underlying parameters that
successfully satisfies the EDM bounds is plotted versus $\thm$. Points
which do not survive a 1\% change are not plotted in the last panel of
Fig.~\ref{fig:mssm151}, although they are plotted in all the other
figures.  It is clear that the large $\thm$ solutions require more
tuning that the low $\thm$ solutions.  Overall, we find that $\{27\%,
46\%, 74\%\}$ of the (large plus small $\thm$) solutions require worse
than $\{1\%,2\%,5\%\}$ tuning (corresponding to \{the unplotted
points, points at 0.01, points at 0.02\} in the last panel of
Fig.~\ref{fig:mssm151}).  The tuning level of the solutions can also
be seen in the other panels of Figs.~\ref{fig:mssm151}, where the open
circles have tuning worse than at the level of $1\%$.

\begin{figure}[b]
\begin{center}
\epsfig{file=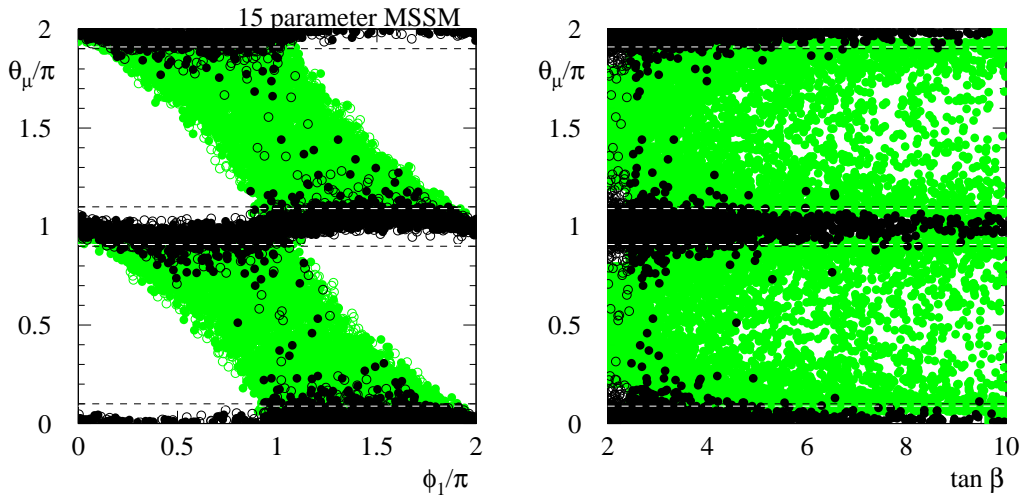,width=14cm} 
\end{center}
\vspace*{-8mm}
 \caption{A scan of the 15-parameter MSSM imposing only the electron
    and neutron EDM constraints.   Light (green) points are forbidden by the $^{199}$Hg EDM
    constraints, while the dark (black)  points satisfy the $^{199}$Hg
    EDM constraints.  The open circles have $m_h<113\gev$.}
    \label{fig:hg}
\end{figure}

We emphasize the importance of having a third independent constraint
on the phases, from the $^{199}$Hg EDM.  Fig.~\ref{fig:hg} shows the
result of a smaller scan of the 15-parameter space in which we only
impose the limits from the electron and neutron EDMs and plot
$\thm$~vs.~$\tb$ for the resulting set.  The light (green) points are
those which would be forbidden by the further imposition of the
$^{199}$Hg EDM constraints, while the dark (black) points are those
which satisfy the $^{199}$Hg EDM constraints. The open circles yield
$m_h<113\gev$.  We find that of the 174,000 points satisfying the
neutron and electron EDM constraints (of which only 50,000 are
plotted), only 4700, or 2.7\%, additionally satisfy the $^{199}$Hg EDM
constraint.  The remaining points at large $\thm$ typically have a
very heavy squark ($>800\gev$) or two heavy charginos ($>500\gev$).
Clearly, ignoring~\cite{choietal} the $^{199}$Hg EDM constraints allows
many configurations that are experimentally forbidden, particularly
for low to moderate values of the masses, and for larger values of
$\tb$ combined with large phases, where the regions satisfying the
electron and neutron EDM constraints are small.  Given the significant
effect of the mercury constraint, an improved calculation of the
strength of the $T$-odd nuclear forces \cite{KKY} and Schiff moment of
the mercury nucleus \cite{KSF} will likewise be important.

\begin{figure}[t]
  \begin{center}
\epsfig{file=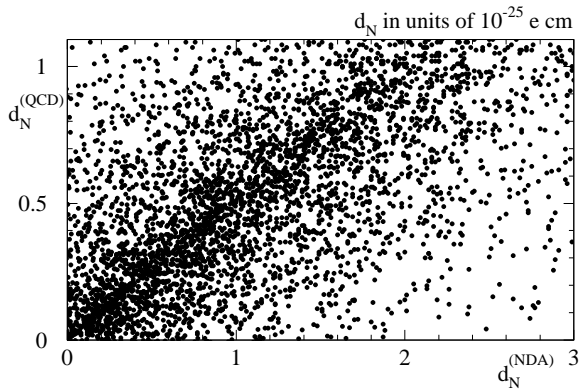,width=8cm}  
  \end{center}
\vspace*{-8mm}
    \caption{Computed values for the neutron EDM, using QCD sum rules
      (y-axis) and na\"{\i}ve dimensional analysis (x-axis), in units
      of $10^{-25} e\cdot {\rm cm}.$\label{fig:nedms}}    
\end{figure}

We also emphasize the significance of the recent improved calculation
of the neutron EDM \cite{PRtheta,PRnedm}.  In Fig.~\ref{fig:nedms} we
display the values of the neutron EDM as computed using QCD sum rules
in \cite{PRtheta,PRnedm} against those estimated using na\"{\i}ve
dimensional analysis (NDA), as in \cite{IN}.  Here we have plotted just
those sets for which the neutron EDM constraint is satisfied
according to either one or both of the calculational methods.  We see
that NDA typically overestimates the neutron EDM by roughly a factor of two,
although for solutions near regions of cancellations, the discrepancy
between the two can be much greater.  

\subsection{23-parameter MSSM}

\begin{figure}[hbtp]
\begin{center}
\epsfig{file=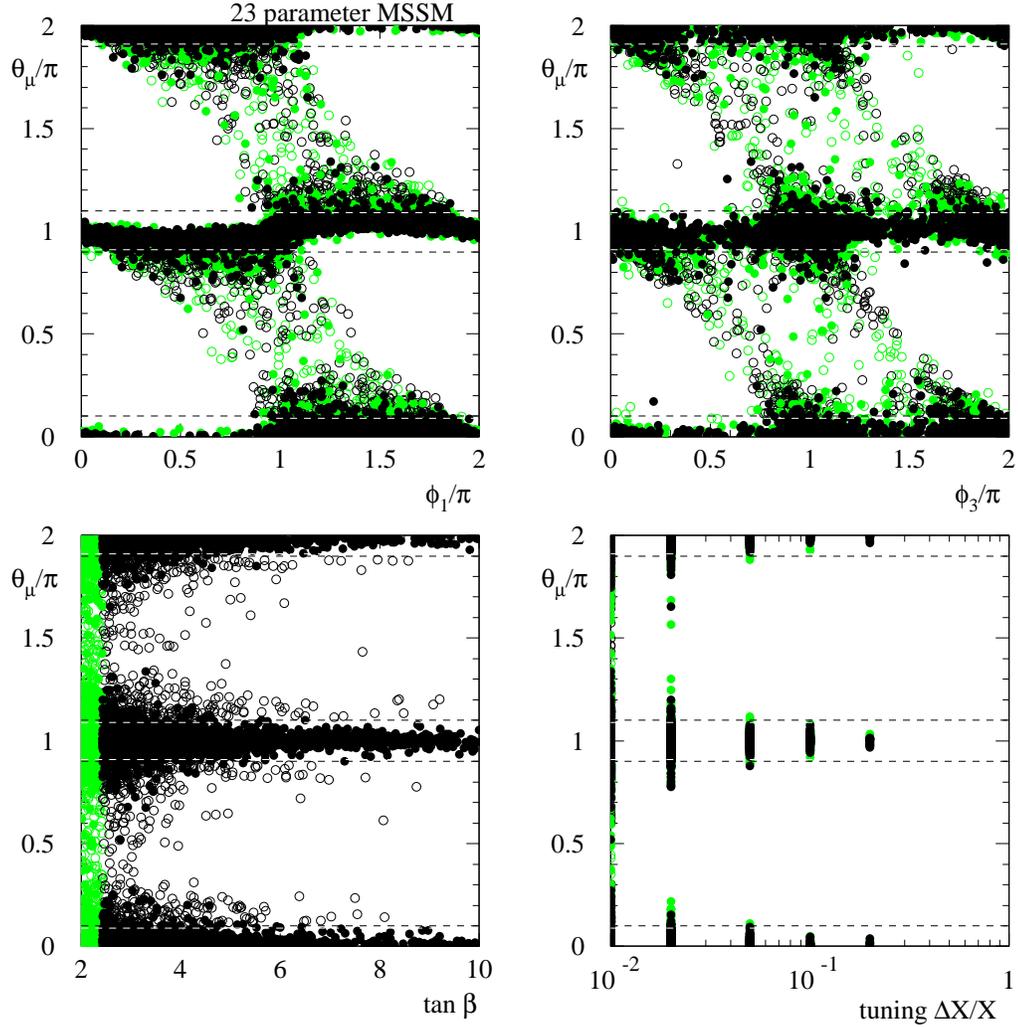,width=14cm} 
\end{center}
\vspace*{-8mm}
  \caption{Parameter sets in the 23-parameter MSSM satisfying the
    experimental limits on the electron, neutron and $^{199}$Hg
    EDMs.  Open circles suffer from parameter
    tuning $\Delta X/X$ worse than 1\% (see
    the text).  Light (green) dots correspond to configurations with a
    light Higgs $m_h<113\gev$.  The tuning parameter $\Delta X/X$ is
    defined in the text, and corresponds to the maximum variation that
    the point survives..}
    \label{fig:mssm23}
\end{figure}

We lastly consider a very general model with 23 free parameters.  We
now allow the left and right sfermion masses to vary independently,
and we take independent stop masses in computing the neutron EDM.  We
also take the phase and magnitude of all the $A_i, i=e,d,u,t$ as independent, giving a
total of 7 phases, 15 mass parameters and $\tb$ as the free parameters
of the model.  We perform a Monte Carlo scan of roughly 600 million
parameter sets, over the following ranges:
\begin{center} \begin{tabular}{rcl}
 0    $\;\;\le$ &$\thm, \theta_{M_i},\theta_{A_{\SP i}}$ &$\le\;\;$   $2\pi$\\
100 {\rm GeV}    $\;\;\le$& $\mu$&   $\le\;\;$ 1 {\rm TeV}\\
100 {\rm GeV}    $\;\;\le$& $2M_1, M_2,M_3$&   $\le\;\;$ 1 {\rm TeV}\\
0 {\rm GeV}    $\;\;\le$& $|A_i|$ &  $\le\;\;$ 1 {\rm TeV}\\
0 {\rm GeV}    $\;\;\le$& $m_{\tilde \ell_{L}}$,$m_{\tilde
  e_{R}}$&  $\le\;\;$ 1 {\rm TeV}\\
0 {\rm GeV}    $\;\;\le$& $m_{\tilde q_{L}}$,$m_{\tilde u_{R}}$,$m_{\tilde d_{R}}$&  $\le\;\;$ 1 {\rm TeV}\\
0 {\rm GeV}    $\;\;\le$& $m_{\tilde q^3_{L}}$,$m_{\tilde t_{R}}$&  $\le\;\;$ 1 {\rm TeV}\\
2    $\;\;\le$&$\tb$&  $\le\;\;$ 10
\end{tabular} \end{center}
In Fig.~\ref{fig:mssm23} we display the 10,000 parameter sets
satisfying the EDM constraints (\ref{eq:elim})-(\ref{eq:hglim}).  The
results are very similar to those in the case of the 15-parameter MSSM.

\section{Mass and Cross Section Measurements}\label{sec:measure}

We now turn to the actual
determination of the phase parameters at a future Linear
Collider~\cite{lincol}. In an earlier work~\cite{mad} it was shown
that large phase values, as well as the real parameters in the
neutralino/chargino sector, can be extracted to high accuracy from
measured masses and cross sections, using a global fit. Their
extraction from a much reduced set of observables, such as the
neutralino and chargino masses alone, appears to be impossible, due to
experimental uncertainties propagating from the measurement of masses
and cross sections into the fitted phase parameters. Only if the
uncertainty is considerably smaller than the actual phase value, do we
regard the phase as being observable.

\begin{figure}[b] 
\begin{center}
\includegraphics[width=16.5cm]{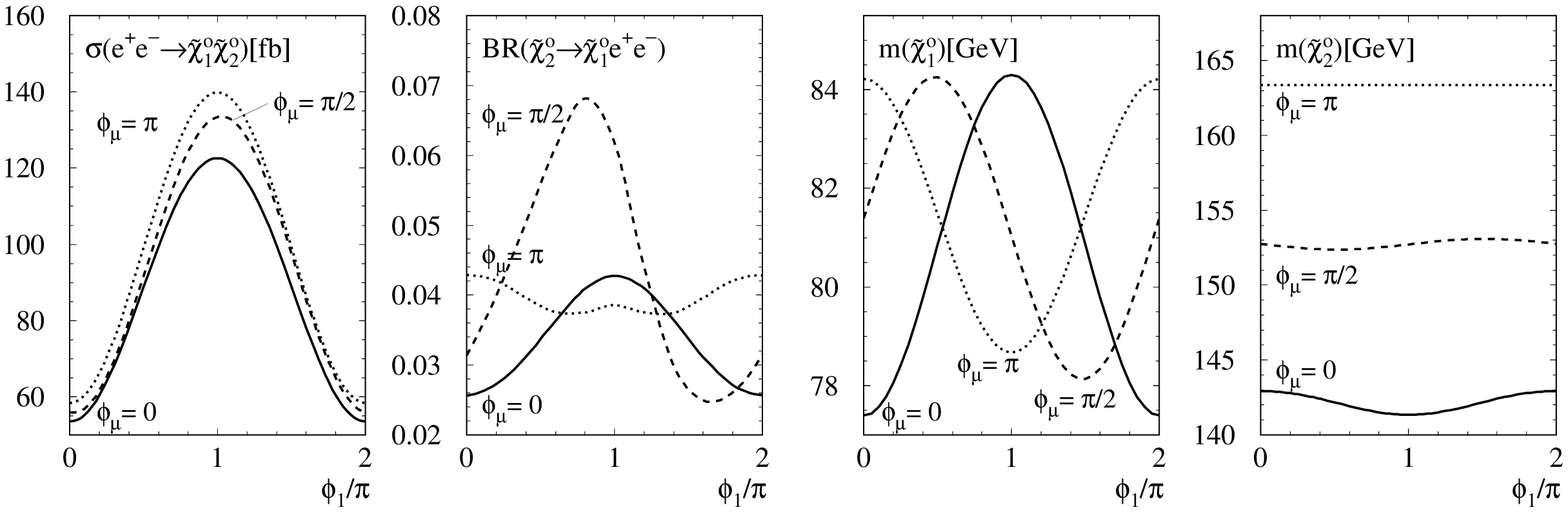}
\end{center} 
\vspace*{-8mm}
\caption[]{The masses and cross sections at a $400 \gev$ Linear 
           Collider for the scenarios discussed in the text. 
          \label{fig:cxn} }
\end{figure}

\subsection{Complete Set of Observables} 

The central scenario in our analysis is a generic set of MSSM
parameters derived from the unified model parameters $m_{1/2}=200
\gev$, $m_0=100 \gev$, $A_0=0$, $\tan\beta=4$, and $\mu>0$.  The
neutralino/chargino mass parameters are $|M_1|=83 \gev, M_2=165 \gev$
and $\mu=310 \gev$, and the corresponding slepton masses are
$m_{\tilde{e}_L}=180 \gev$, $m_{\tilde{e}_R}= 132 \gev$ and
$m_{\tilde{\nu}}=166 \gev$.  The masses of the neutralinos and
charginos are given in Table~\ref{tab:cxn}. For the phase values
$\phi_1=\theta_\mu=\pi/10,$ this scenario was investigated in our
previous analysis: the error on the fitted mass parameters $|M_1|,
M_2, |\mu|$ is smaller than $1 \gev$, and the error on the extraction
of $\tan\beta$ is less than $10 \%$. The RMS of the phase
determination depends on the energy and the luminosity of the
collider.  For a ($500 \gev, 500 \ifb$) machine we obtain $\phi_1/\pi
= 0.1 \pm 0.03$ and $\theta_\mu/\pi = 0.1 \pm 0.05$, whereas for ($1
\tev, 1000 \ifb$), the errors are $\phi_1/\pi=0.1 \pm 0.05$ and
$\theta_\mu/\pi=0.1 \pm 0.06$\footnote{For both colliders the masses
  are determined in a threshold scan. For higher energy and fixed
  masses the cross sections decrease. Increasing the luminosity does
  not compensate for the increase in energy in this case.}. The
$t$-channel slepton masses are assumed to be measured in threshold
scans, and the propagation of the uncertainty of their mass
measurements into the phases can be neglected. We find that the
statistical errors on the phase determination $\Delta \phi_1 \sim 0.03
\pi$ and $\Delta \theta_\mu \sim 0.05 \pi$ are essentially independent
of the central values of the phases. This implies that even with a
maximal set of observables, phase values smaller than $\pi/10$ are
hidden by experimental errors and are therefore unobservable: CP
phases of that size could, from a collider phenomenology point of
view, as well be zero. \medskip

A na\"{\i}ve view of the inversion problem is given by the graphs in
Fig.~\ref{fig:cxn}. We fix the parameters in the neutralino and 
chargino mass matrices and plot the production cross section
$\sigma(e^+e^-\to\nz{1}\nz{2})$ and the branching fraction 
BR~$(\nz{2}\to\nz{1}e^+e^-)$ as a function of the phase $\phi_1$.
The dependence on $\phi_1$ is strong 
in both observables and an analysis could be
straightforward. However, the plots also show that the physical masses
in the process are not constant with varying phases, either. The main
problem of the analysis therefore becomes the separation of direct
effects of the phase and indirect effects, where the phase changes the
physical neutralino masses and these affect the cross sections and
branching fractions.

Hence, a key ingredient of the full analysis~\cite{mad} as well as of
the one presented in this paper is the measurement of masses through
threshold scans~\cite{blair}. We use the error estimates for an mSUGRA
scenario with $m_0= 100\gev$, $m_{1/2}=200 \gev$ and $\tan\beta=4$.
They are $0.05/0.07/0.3/0.6 \gev$ for the neutralino masses and
$0.035/0.25 \gev$ for the chargino masses~\cite{blair}. We
furthermore assume the uncertainty of cross section measurements to be
purely statistical. This limits the pull of small cross sections. The
invisible cross section for the production of two lightest neutralinos
(LSPs) is not part of our sample. For a high energy collider, the full
analysis includes twelve cross section and six mass observables,
determining six model parameters: $|M_1|, M_2, |\mu|, \tan\beta,
\phi_1, \theta_\mu$.

\begin{table}[b] 
\begin{center}
\begin{tabular}{|c|r||c|r|r|r|}
 \hline
  & $m_{\tilde{\chi}}$~[GeV]  &
  & $\sigma_{\rm tot}$~[fb], $\tilde{\ell}$ light & 
    $\sigma_{\rm tot}$~[fb], $\tilde{\ell}$ heavy & 
    $\sigma_{\rm tot}$~[fb], $\tilde{\ell}$ decoupled \\[1mm]
 \hline \hline
 $\nz{1}$         &  77.3    & 
 ($\nz{1} \nz{1})$ &  320.0   &  186.6   &  0.09  \\
 $\nz{2}$         &  142.6   & 
  $\nz{1} \nz{2}$  &  87.8    &  55.3    &  0.28  \\
 $\nz{3}$         &  315.5   & 
  $\nz{1} \nz{3}$  &  6.4     &  3.8     &  1.7   \\
 $\nz{4}$         &  343.0   & 
  $\nz{2} \nz{2}$  &  107.8   &  64.5    &  0.16  \\
 $\cp{1}$         &  140.7   & 
  $\cp{1} \cm{1}$  &  806.9   &  570.8   &  854.6 \\
 $\cp{2}$         &  431.6   & 
                   &          &          &        \\
 \hline
\end{tabular} \vspace*{0.3cm}
\caption[]{\label{tab:cxn}
Neutralino/chargino masses and cross sections 
at a 400~TeV Linear Collider
for the three scenarios under consideration. The production cross
section of two LSPs is not part of the set of observables.}
\end{center}
\end{table}

\subsection{Inclusion of the EDM limits}

To cover a large range of slepton masses, we consider two
modifications of the central scenario: (1) the first generation
sleptons are still light but just escape detection at a low energy
collider. The so increased slepton masses correspond to having their
$m_0 = 200 \gev, m_{1/2} = 200 \gev$, {\it i.e.}  $m_{\tilde{e}_L}=250
\gev$, $m_{\tilde{e}_R}=218 \gev$ and $m_{\tilde{\nu}}=240 \gev$. (2)
All sleptons decouple from the theory entirely, which corresponds to
masses above ${\cal O}(1 \tev)$. The cross sections accessible at a
$400 \gev$ collider are given in Table~\ref{tab:cxn} for all three
sets of lepton masses. This collider energy enables us to observe one
higgsino type neutralino directly, whereas the other two higgsino
states are kinematically inaccessible. The cross sections are
dominated by the $t$-channel process due to the gaugino nature of the
light neutralino. In contrast, the production of two light charginos
involves both diagrams: for light sleptons, the $t$-channel graph is
large whereas in the decoupling limit the $s$-channel contribution
dominates; for intermediate slepton masses destructive interference
reduces the cross section by almost a factor of two.  Still, the
chargino cross section is by far the largest. If the lightest chargino
mass is known through a threshold scan, this signature should serve to
determine the mass of the sneutrino with high
precision~\cite{gudi}.\medskip

Using the results of the EDM analysis allows us to modify the full
analysis described in the previous section: we know that the phase
$\theta_\mu$ has to be smaller than $\pi/10$, after taking into account
three conditions: (1) the experimental limits on the electron,
neutron and mercury EDM have to be respected simultaneously; (2) the
degree of fine tuning is limited by requiring that the solutions be
stable with respect to changes of $\Delta X/X=1\%$ in all model
parameters; (3) a minimal set of final states is produced at a Linear
Collider with a fixed design energy, {\it e.g.} $400 \gev$. We will
show that this minimal set of observables 
has to include a handle on one higgsino
component, to allow for a determination of $\mu$. Under these
assumptions, we try to determine CP-violating phases at a comparably
low energy collider, which would not cover all neutralino/chargino and
slepton thresholds. The reduced set of observables is supplemented by
EDM constraints: $\tan \beta $ has to be small ($\la 10$), and may
be known from the Higgs sector~\cite{tanbeta} if the
Higgs bosons are sufficiently light. Since $\theta_\mu$
turns out to be constrained to be smaller than its minimum visible
value $\pi/10$~\cite{mad}, we fix it to zero. The phase
$\phi_1$ can be treated as independent of all other parameters in the
neutralino and chargino mass matrices, since parameters correlated to
$\phi_1$ by the EDM constraints, like $\phi_3$ or $\phi_A$, do not
appear in this sector.  \smallskip

To determine the phases we modify our full analysis: the set of
observables is first reduced to the three lightest neutralinos and the
lightest chargino.  Producing them in pairs yields four cross
sections, {\it i.e.} eight independent observables.  In the fit, we
determine the real parameters $|M_1|, M_2, \mu, \phi_1$ as well as two
$t$-channel selectron masses. For a low energy collider, the selectron
masses cannot be expected to be measured in threshold scans, and this
leaves us with altogether six unknown model parameters.  Using $SU(2)$
symmetry the sneutrino mass can be related to the left handed
selectron mass:
\begin{equation}
m_{\tilde{\nu}}^2 = m_{\tilde{e}_L}^2 
                  + m_W^2 \cos^2\theta_w \cos 2 \beta~.
\end{equation}
In contrast to the complete parameter fit, the extraction is now
limited by the number of fitted parameters, and the $\chi^2$
distribution of the best fit might be flat in certain parameters. This
makes it technically difficult to add $\tan\beta$ to the set of fitted
parameters, and we have to rely on a measurement in the Higgs sector.
However, in principle one can extend the number of observables by
non-trivial distributions, asymmetries or additional cross sections,
and one can choose a fitting algorithm better suited for the problem
to include $\tan\beta$ in the fit~\cite{cpeven}.

\subsection{Statistical Uncertainties} 

The inclusion of experimental errors follows the same path as in the
previous analysis~\cite{mad}: we assume Gaussian errors for the
measured masses and cross sections and define smeared
pseudo-measurements. First we define a set of `true' model parameters
$|M_1|, M_2...$ They predict a set of `true' observables (masses and
cross sections), all of which are assumed to have a Gaussian error
distribution with a known width~\cite{blair}.  Using these
distributions, we randomly vary 10000 pseudo-measurements of the set
of observables. These sets become slightly inconsistent, but using a
global fit we can extract the central value of every model parameter
and obtain a distribution for each of them\footnote{Extracting these
  values from the low energy parameters without the smearing would
  merely serve as a check of the fitting program.}. In the fit we
minimize $\chi^2 = \sum_i (x_{\rm reconstr,i}-x_{\rm meas,i})^2/e_{\rm
  i}^2$, where $x_{\rm reconstr}$ are the reconstructed observables
and $x_{\rm meas}$ are the smeared pseudo-measurements. The error on
the cross sections is given as a function of the luminosity by 
$e_{\rm i}
= \sigma_{\rm i} /(\epsilon {\cal L})$, where the efficiency is
assumed to be $\epsilon = 10\%$.  If the central value of the
distribution for a given parameter agrees with its `true' value, the
statistical treatment is justified, and the width of the distribution
describes the migration of observational errors into the model
parameters. The final curve does not necessarily have to be Gaussian,
since correlations, together with the range of the fit, can alter the
shape of the curve. Hence, we quote the RMS value instead of the
standard deviation of a fitted Gaussian distribution. We note that for
some scenarios there exist several $\chi^2$ minima in different
parameter regions. This is a technical complication, and the minima
should be distinguishable through the actual values of $\chi^2$.  In
these cases, we limit our range of fitted parameters, making sure that
the range is much bigger than the distribution we finally obtain.
\medskip

\begin{figure}[b] 
\begin{center}
\includegraphics[width=11.2cm]{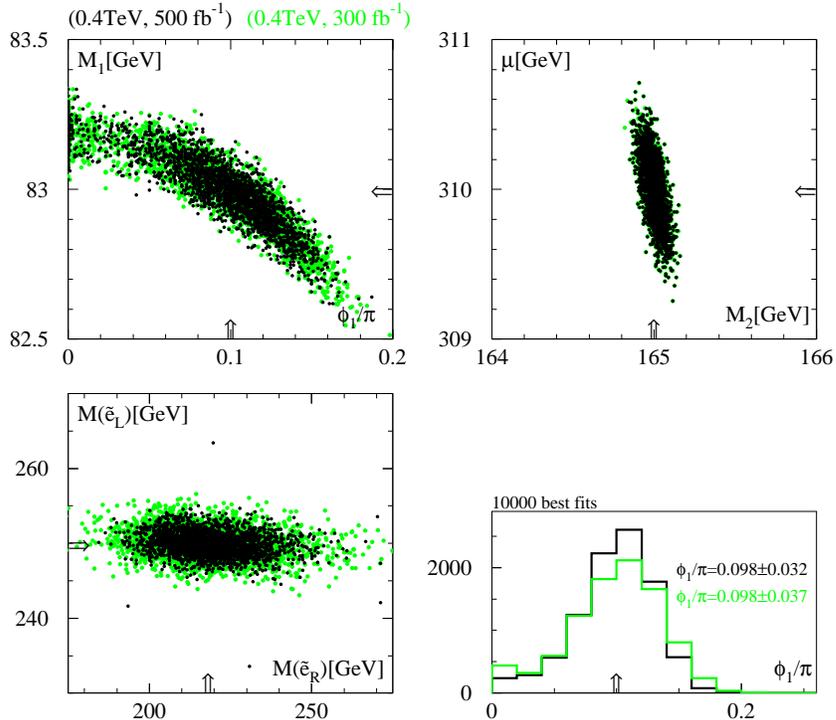} 
\end{center} 
\vspace*{-8mm}
\caption[]{Parameters obtained from best fits to 
  10000 pseudo-measurements. The central scenario is indicated by
  arrows on the axes. Only one Higgsino $\nz{3}$ is assumed to be
  observed. The black and grey (green) points correspond to two
  different collider designs. \label{fig:bestfit} }
\end{figure}

The distribution of best fits to a set of 10000 pseudo-measurements is
given in Fig.~\ref{fig:bestfit}. In contrast to the earlier analysis
we  assume $\theta_\mu=0$ and rely on a known value of $\tan\beta$.
The sleptons are kinematically inaccessible, 
and their masses have to be indirectly determined
from the cross section measurement. Comparing the errors on the three
mass parameters $|M_1|, M_2, \mu$ we notice the striking accuracy of
the measurement of $M_2$. This reflects the small number of unknown
parameters in the chargino mass matrix and in the $\cp{1} \cm{1}$
production cross section: the phase $\theta_\mu$ and $\tan\beta$ are
both fixed, and only $M_2$ and $\mu$ and the sneutrino mass have to be
determined from the chargino mass and cross section. However, if one
is able to determine the cross sections for polarized electrons and
positrons and separate the $s$ and $t$-channel contributions, this
accuracy will improve even further. On the other hand, if one is
forced to determine $\tan\beta$ from the neutralino/chargino sector,
the error on $M_2$ will be ${\cal O}(1 \gev)$ again. In contrast to
the full analysis, the fitted values of $|M_1|$ and $\phi_1$ are now
correlated.  Correlations like that are a general feature of a smaller
set of observables and a smaller dimension of the fit. The error on
the determination of the slepton masses is not symmetric: the left
selectron mass is tied to the sneutrino mass and therefore extracted
from the well determined chargino sector. The right selectron mass, in
contrast, has to be determined from the neutral gaugino cross
sections, which are smaller by almost one order of magnitude.

The distribution of best fits for $\phi_1$ exhibits a similar shape to
the result of the full analysis. For large luminosity of $500 \ifb$ it
approaches a Gaussian with a RMS of 0.032 and a fitted standard
deviation of 0.028. At smaller luminosity, this distribution broadens,
{\it e.g.} to a RMS value of 0.037 for $300 \ifb$, as given in
Table~\ref{tab:fit}. From the cross sections in Table~\ref{tab:cxn}
one can see that there is a minimal luminosity at which the number of
events in $(\nz{1} \nz{3})$ becomes marginal. In this case there will
be hardly any information on $\mu$ left in the sample.  The same
problem arises once the energy is too small to produce a state with
significant higgsino content. If the higgsinos are almost entirely
decoupled, the uncertainty on the real mass parameters becomes too
large to extract the phases. Technically, the correct minimum in
$\chi^2$ will not be found then by the fitting procedure. In
Figure~\ref{fig:cxn} we see that for many observables the point
$\phi_1=0$ is extremal, {\it i.e.} a considerable number of best fits
will give this result. This occurs for example in the case of low
luminosity, Fig.~\ref{fig:bestfit}.  On the other hand we notice that
in the set of observables under consideration, the gaugino sector is
still over-determined: except for the less important slepton masses
the $\nz{2} \nz{2}$ production cross section carries similar
information as the much larger $\cp{1} \cm{1}$ cross section.
Removing the $\nz{2} \nz{2}$ cross section from the sample increases
the RMS value given in Table~\ref{tab:fit} to 0.036 ; however, in most
models $\nz{2}\nz{2}$ will be visible at any collider that produces
pairs of light charginos.

\subsection{Systematic Uncertainties} 

\begin{figure}[b] 
\begin{center}
\includegraphics[width=11.2cm]{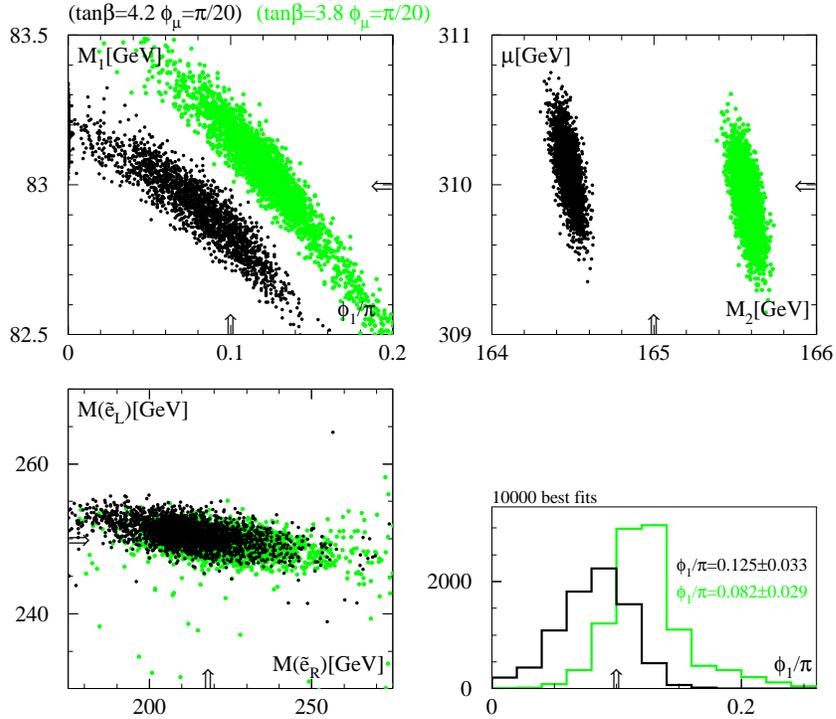}
\end{center} 
\vspace*{-8mm}
\caption[]{The effect of wrongly assumed values for 
           $\tan\beta$ on the extraction of the model 
           parameters. The `true' value for $\tan\beta$
           is 4. \label{fig:wrongfit} }
\end{figure}

As mentioned above the errors derived above do not take into account
any systematical errors from using a wrong value {\it e.g.} of
$\tan\beta$. Neither do they include an estimate what would happen if
$\theta_\mu$ were actually small but non-zero. Both effects would lead
to systematic errors and thereby to wrong central values of the 10000
fits. In Table~\ref{tab:fit} we show that fitting a set of parameters
with a `true' value $\theta_\mu=\pi/20$, but assuming $\theta_\mu=0$
hardly affects the fit.  This observation serves as a consistency
check of the ansatz, namely that a small $\theta_\mu$ has no effect on
the extraction of $\phi_1$ and certainly not on the dominating real
parameters in the mass matrices.  Analyzing the set of observables
with a wrongly assumed values of $\tan\beta=3.8$ or $\tan\beta=4.2$
instead of a `true' value of 4.0 leads to the systematic error shown
in Fig~\ref{fig:wrongfit}: the spread of the best fits in the mass
parameters as well as in the phase is very similar to the case where
both parameters, $\theta_\mu$ and $\tan\beta$ are input correctly, but
the central values are shifted.  The central values for wrongly assumed
values $\tan\beta=3.8$ and $4.2$ become $\phi_1/\pi=0.125 \pm 0.033$
and $\phi_1/\pi=0.082 \pm 0.029$, {\it i.e.} they are shifted by half
the RMS value.

To estimate the dependence of the analysis on the details of the
underlying model, we decouple the sleptons from all cross sections.
From the cross sections in Table~\ref{tab:cxn}, we see that the real
mass parameters dominated by the physical chargino/neutralino masses
and the chargino cross section should be extracted as precisely as in
the case of light sleptons.  However, the determination of the phase
$\phi_1$ relies on the measurement of the $\nz{1} \nz{2}$ and the
$\nz{1} \nz{3}$ cross section, where the $Z$ boson in the $s$-channel
production process for neutralinos only couples to the higgsino
fraction of the physical states. Since in our central scenario the
light neutralinos are mainly gauginos, both of these cross sections
decrease with increasing slepton masses. This leads to a RMS value of
$0.044$ of the $\phi_1$ distribution, as seen in Table~\ref{tab:fit}.

An important feature of the complete analysis~\cite{mad} is that the
error on the $\phi_1$ measurement is independent of the actual value
of $\phi_1$. This allows us to derive lower limits on the size of
observable phases, which are $\phi_1 \ga \pi/10$ and $\theta_\mu
\ga \pi/10$ for the complete analysis. As presented in
Table~\ref{tab:fit}, we obtain RMS=$0.030$ for a central value of
$\phi_1=\pi/2$.  In Figure~\ref{fig:cxn}, the cross section, as well as
the branching fraction, tends to vary strongly around phase values of
$\pi/2$, while they become flat and extremal for positive or negative
real values of $M_1$. This is reflected in the slightly smaller RMS
value for $\phi_1=\pi/2$ as compared to $\phi_1=\pi/10$.

From the cross sections in Table~\ref{tab:cxn} it is obvious that the
determination of $\phi_1$ from the given set of cross sections
improves, once $\mu$ takes a value closer to the gaugino mass
parameters. Reducing it by a factor of two to $\mu=150 \gev$ leads to
a much lighter higgsino with $\mne{3}=158 \gev$ and an increased cross
section for $\nz{1}\nz{3}$ production. The changes in the masses of
the gaugino dominated light neutralinos are small, but the error on the
$\phi_1$ measurement improves to 0.014. \smallskip

Although it is hard to accommodate with the EDM constraints, we
investigate a scenario with large $\tan\beta=30$: from the full
analysis we expect the determination of $\tan\beta$ to be less precise
than for smaller values.  Since all production cross sections are
smaller by up to $50 \%$, the extraction of the slepton masses becomes
increasingly difficult. From the analysis for small values of
$\tan\beta$, however, we know that the error on the determination if
the right selectron mass is larger than the mass difference between
the two selectrons. One way of improving the fit for large $\tan\beta$
is therefore to assume that the selectrons are mass degenerate. With
this assumption we are able to
 determine the phase $\phi_1=0.11 \pm
0.03$.

\subsection{Minimal Set of Observables} 

\begin{figure}[b] 
\begin{center}
\includegraphics[width=16.0cm]{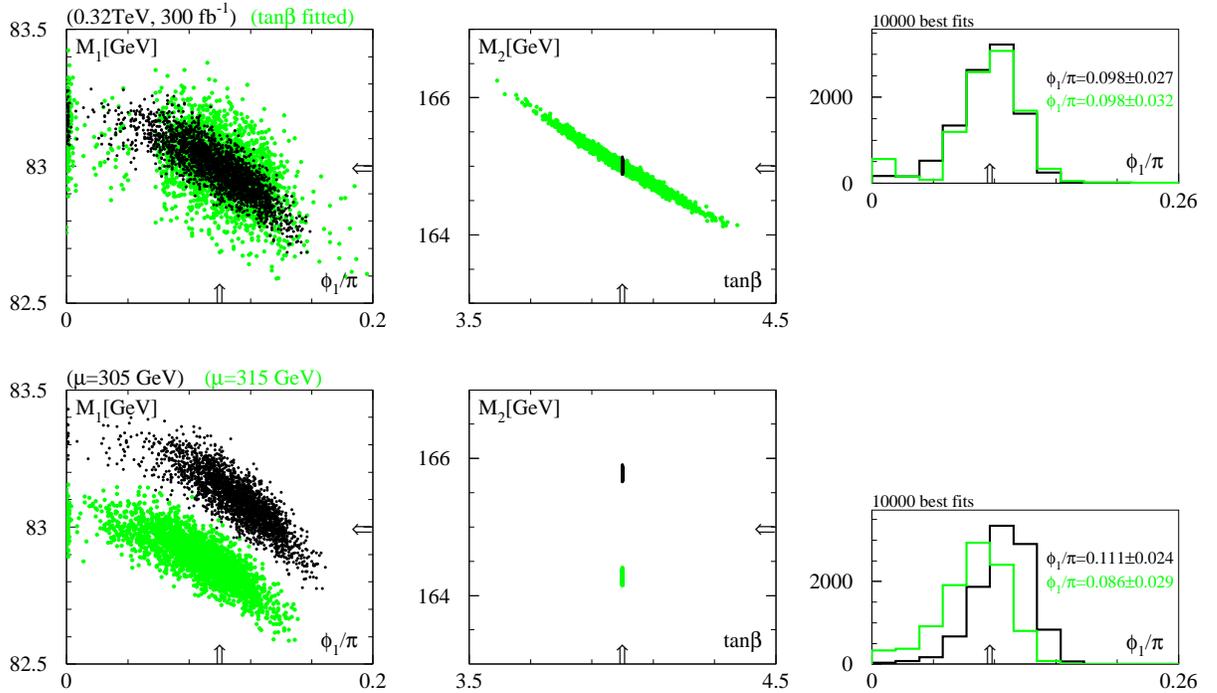}
\end{center} 
\vspace*{-8mm}
\caption[]{Top: Fit results for a known value of $\mu=310 \gev$.
           The grey (green) points correspond to $\tan\beta$ being 
           added to the fitted parameters. Bottom: Systematic effect 
           from a wrong input of $|\mu|$ for known $\tan\beta$.
           \label{fig:minimal} }
\end{figure}

Since the EDM constraints on $\theta_\mu \la \pi/10$ require the
chargino sector to be essentially CP-conserving, we can define a
minimal set of observables sufficient to extract the phases: in the 
CP-conserving case one could be able to extract $|\mu|$ sufficiently
precisely from a set of the mass, cross section and asymmetries in
$\cm{1}\cp{1}$ production analytically~\cite{cpeven,jan}.
  We use this
result to investigate the phases in the neutral gaugino sector,
including the masses and production cross sections of $\nz{1}$ and
$\nz{2}$. In our central scenario (Table~\ref{tab:cxn})
this requires a minimum collider energy of $320 \gev$. 
To the set of three
cross sections and three masses we fit the neutralino parameters
$|M_1|, M_2, \phi_1$, for given $|\mu|$, and obtain an error on the
determination of $\phi_1$ of $0.027$, {\it i.e.}  similar to case
including one higgsino. Adding $\tan \beta$ to the fitted parameters
increases the error on $\phi_1$ to RMS$=0.032$, as can be seen in
Fig.~\ref{fig:minimal}. As shown before, the inclusion of slepton
masses would not change this result significantly. It only requires a
larger set of observables. A wrong measurement of $|\mu|$, however,
would lead to a systematic error of the phase determination.  We
estimate this by assuming a wrong extraction of $|\mu|=305 \gev$ and
$315 \gev$ for a `true' value $|\mu|=310 \gev$. In
Fig.~\ref{fig:minimal} the extracted values of $|M_1|$ and $M_2$ are
shifted systematically, as is the central value of the 10000
pseudo-measurements of $\phi_1$. But the shift for a $5 \gev$
mismeasurement of $|\mu|$ is only by half the statistical error of the
phase measurement. We therefore conclude that it is possible to
extract the phase $\phi_1$ from a minimal set of parameters, if the
error on the determination of $|\mu|$ does not exceed a few
percent~\cite{jan}.

\begin{table}[htb] 
\begin{center}
\begin{tabular}{|cl|c|c|}
\hline
&  & central $\phi_1/\pi$ & RMS \\
\hline \hline
(1) & complete analysis $(500 \gev, 500 \ifb)$~\cite{mad}  & 
   0.097 & 0.030 \\  
(2) & one phase $\phi_1$, free sleptons, fixed $\tan\beta\;$
               $(500 \gev, 500 \ifb)$  &
   0.099 & 0.034 \\
(3) & as (2), but only one higgsino $\tilde{\chi}^0_3$ $(400 \gev, 500 \ifb)$&
   0.098 & 0.032 \\
    & as (3), but $(300 \ifb)$&
   0.098 & 0.037 \\
    & as (3), but without $\nz{2}\nz{2}$ production &
   0.098 & 0.036 \\
    & as (3), but `true' $\theta_\mu=\pi/20$ &
   0.100 & 0.032 \\
    & as (3), but `true' $\theta_\mu=\pi/20$ and wrongly assumed  $\tan\beta=3.8$ &
   0.125 & 0.033 \\
    & as (3), but `true' $\theta_\mu=\pi/20$ and wrongly assumed  $\tan\beta=4.2$ &
   0.082 & 0.029 \\
    & as (3), but sleptons decoupled &
   0.104 & 0.044 \\
    & as (3), but large phase value $\phi_1 = \pi/2$ &
   0.501 & 0.030 \\
    & as (3), but light higgsinos $\mu=150 {\rm \, GeV}\;$
              $(300 \gev, 500 \ifb)$ &
   0.100 & 0.014 \\
(4) & only gauginos $(320 \gev, 300 \ifb)$&
   0.098 & 0.027 \\
    & as (4), but $500 \ifb$&
   0.099 & 0.021 \\
    & as (4), but $\tan \beta$ fitted&
   0.098 & 0.032 \\
    & as (4), but wrongly assumed $\mu=305 \gev$&
   0.111 & 0.024 \\
    & as (4), but wrongly assumed $\mu=315 \gev$&
   0.086 & 0.029 \\
\hline
\end{tabular} \vspace*{0.3cm}
\caption[]{\label{tab:fit}
Central and RMS values for the distribution of
$\phi_1$ values, fitted from the 10000 
pseudo-experiments, with a ``true'' value of $\phi_1=0.1\pi$ (except
where indicated).}
\end{center}
\end{table}

\section{CP-Odd Variables and Asymmetries}

To unambiguously identify an effect of CP violation, 
one needs to construct a ``CP-odd variable'', 
whose expectation value vanishes if CP is conserved.
At an $e^+e^-$ linear collider, the 
initial state can be made a CP eigenstate, given
the CP transformation relation
\begin{equation}
e^-( \sigma_1,\vec p)\ e^+( \sigma_2,-\vec p) \Rightarrow 
e^-(-\sigma_2,\vec p)\ e^+(-\sigma_1,-\vec p),
\label{cp}
\end{equation}
where $\sigma_i$ is the fermion helicity.
We consider a specific process
\begin{equation}
e^-(p_1)e^+(p_2) \to \nz{1} \nz{2} \to \ell_1^-(q_1) \ell_2^+(q_2) 
\nz{1} \nz{1}.
\label{chi12}
\end{equation}
Denote a helicity matrix element by
${\cal M}_{\sigma_1\sigma_2}(\vec q_1,\vec q_2)$
where $\sigma_1\ (\sigma_2)$ is the helicity of the
initial state electron (positron), which coincides with the
longitudinal beam polarization; $\vec q_1\ (\vec q_2)$ denotes 
the momentum of the final state fermion (anti-fermion).
For the process of Eq.~(\ref{chi12}), only two combinations of
the helicity amplitude ${\cal M}_{-+},\ {\cal M}_{+-}$ contribute.
It is easy to show that under CP transformation,
\begin{eqnarray}
\label{pm}
{\cal M}_{-+}( {\vec q_1}, {\vec q_2}) \Rightarrow
  {\cal M}_{-+}(-{\vec q_2},-{\vec q_1}),
\end{eqnarray}
and ${\cal M}_{+-}$ transforms similarly.
If CP is conserved in the reaction, relation (\ref{pm}) 
becomes an equality. This argument is applicable
for unpolarized or transversely polarized beams as well.

One can construct CP-odd kinematical variables
to test the CP property of the reaction. We consider
the following three angles defined as
\begin{eqnarray}
\label{thetaz}
\cos\theta_+={\vec p_1\cdot \vec q_+ \over |\vec p_1|\ |\vec q_+|},\quad
\cos\theta_-={(\vec p_1\times \vec q_-)\cdot (\vec q_1\times \vec q_2)
\over |\vec p_1\times \vec q_-|\ |\vec q_1\times \vec q_2|},\quad
\cos\theta_\ell={\vec p_1\cdot (\vec q_1\times \vec q_2)
\over |\vec p_1|\ |\vec q_1\times \vec q_2|},
\end{eqnarray}
where $\vec q_+=\vec q_1+\vec q_2$ and  $\vec q_-=\vec q_1-\vec q_2$.
It is easy to verify that all the three variables are CP-odd
under final state CP transformations.
We can then construct ``forward-backward'' asymmetries 
\begin{equation}
{\cal A}^{FB}=\sigma^F-\sigma^B
=\int_{0}^{1}{  d\sigma\over d\cos\theta} d\cos\theta
-\int_{-1}^{0} {d\sigma\over d\cos\theta} d\cos\theta,
\label{generalbf}
\end{equation}
with respect to a CP-odd angle $\theta$.

Of the four Feynman diagrams contributing to process (\ref{chi12}),
there are two diagrams that contain explicit CP-violating phases, one
from $s$-channel $Z$ exchange and one from a $t$-channel selectron
exchange. If $\mu > |M_1|, M_2$ ($\mu < |M_1|, M_2$), then in most
part of the parameter space the contribution from the selectron ($Z$)
exchange diagram is dominant, thus makes the CP asymmetry from the
interference small. We scanned the parameter space in
$(\phi_1,\theta_\mu)$, the asymmetry obtained is typically $0.1-1.5\%$.  
The beam polarizations do not improve the situation significantly.  
As an example, for $|M_1| = 80 \gev$, $M_2 = 200 \gev$,
$\mu = 275 \gev$, $\tan\beta = 4.0$ and $m_{\tilde e_{R}} = 165 \gev$,
while $\phi_1 = 0.90\pi$ and $\theta_\mu = 0.25 \pi$ the asymmetry
from $\cos\theta_\ell$ appears to be about $1.0\%$, 
with an asymmetry rate of 2.7 fb. 
The asymmetries for other variables in Eq.~(\ref{thetaz}) are comparable.
Naively, such an asymmetry at a high luminosity linear collider of
200 fb$^{-1}$ would lead to a 2$\sigma$ statistical effect, as
estimated in \cite{choietal}. However, one would have to keep 
the systematics of the asymmetry measurements well below a percent
level, in order to establish a positive observation. We thus consider
it very challenging to experimentally observe this rather small asymmetry.

\section{Conclusions/Outlook}

We have shown that the current experimental limits on the neutron,
electron and $^{199}$Hg electric dipole moments strongly constrain
general SUSY models with several CP-violating phases, even in the
presence of strong cancellations between the various SUSY
contributions to the EDMs.  Although it is only $\thm$ which is 
typically constrained to be small,   in models in which the phase of $B$
is correlated to the other (large) phases in the model, this
translates into a tight restriction on some combination of (large)
phases, which may not be natural in any given model.
The next year will see significant
improvements in the experimental limit on the electron EDM
\cite{commins}, and work on improving the mercury EDM limits
continues as well.  Further down the road, new experimental
techniques, such as using diatomic molecules to measure the electron
EDM \cite{dbbk}, or studies of the effects of CP violation on other systems, 
such as ($b\bar b$) production\cite{demir},
 will probe additional CP violation in the MSSM to still more sensitive levels. 

These constraints on CP-violating phases are important 
for studies at a future Linear Collider. It was shown that 
one can always extract the phases $\phi_1$ and $\theta_\mu$ from a 
complete set of masses and cross sections for neutralino and 
chargino pair production~\cite{mad}. The new EDM constraints 
essentially require $\theta_\mu$ to be too small to be measured at 
a Linear Collider. In the region where charginos and neutralinos 
are visible at a Linear Collider $\tan\beta$ is preferably 
small, {\sl i.e.} rendering its measurement accessible through Higgs 
production~\cite{tanbeta}. Using the above constraints we extracted 
the finite phase $\phi_1$ from a reduced set of masses and cross 
sections and find that a minimal set of observables is limited by 
the presence of a non-vanishing Higgsino component 
in the final state particles. 
Since a lower energy ($400 \gev$) Linear Collider 
might not be able to produce
slepton pairs we show that the $t$-channel masses in chargino and
neutralino production can be fitted, their uncertainty does not 
interfere with the $\phi_1$ phase measurement. 
Pair production 
of light charginos~\cite{cpeven} might provide us 
with precise indirect information on the size of 
$|\mu|$~\cite{jan}. We showed that in this case that we do not need 
to produce any chargino directly to determine the phase 
$\phi_1$ from a truly minimal set of (gaugino) observables. For this 
set of observables we can even measure $\tan\beta$ and easily 
cross check a $\tan\beta$ measurement from the Higgs sector.

We finally studied CP-odd variables at $e^+e^-$ linear colliders.  We
found that the CP decay asymmetry constructed from the final state
kinematics is typically about $0.1-1.5\%$.  It will be challenging to
experimentally observe this rather small asymmetry.

\vskip 0.5in
\vbox{
\noindent{ {\bf Acknowledgments} } \\
\noindent
We thank the organizers of the ``30 Years of Supersymmetry'' Workshop in
Minneapolis, where some of this work was completed.  We would like
thank Maxim Pospelov for many discussions, along with Norval Fortson,
Mike Romalis, Jan Kalinowski and Hitoshi Murayama.  TF would also like
to thank the TPI in Minneapolis for its gracious hospitality this
fall.  This work was supported in part by DOE grant
DE--FG02--95ER--40896, in part by the University of Wisconsin
Research Committee with funds granted by the Wisconsin Alumni Research
Foundation, and in part by DOE grant DOE-EY-76-02-3071.}

\end{document}